\documentclass{ws-procs975x65}

\def\beq{\begin{equation}}
\def\eeq{\end{equation}}

\def\Msol{\mbox{ }M_{\odot}}

\def\ergs{\mbox{ erg\,s}^{-1}}

\def\amin{^\prime}
\def\adeg{^{\circ}}
\def\nh{N_{\rm H}}
\def\cmmoinsdeux{\mbox{ cm}^{-2}}

\begin{document}

\title{High Mass X-ray Binaries: Progenitors of double neutron star systems}
\author{Sylvain Chaty$^*$}

\address{Laboratoire AIM (UMR 7158 CEA/DSM-CNRS-Universit\'e Paris Diderot), Irfu/Service d'Astrophysique, Centre de Saclay, FR-91191 Gif-sur-Yvette Cedex, France \\
Institut Universitaire de France, 103, boulevard Saint-Michel, FR-75005 Paris, France \\
$^*$E-mail: chaty@cea.fr\\}

\begin{abstract}
In this review I briefly describe the nature of the three kinds of High-Mass X-ray Binaries (HMXBs), accreting through: (i) Be circumstellar disc, (ii) supergiant stellar wind, and (iii) Roche lobe filling supergiants. A previously unknown population of HMXBs hosting supergiant stars has been revealed in the last years, with multi-wavelength campaigns including high energy ({\it INTEGRAL, Swift, XMM, Chandra}) and optical/infrared (mainly ESO) observations. This population is divided between obscured supergiant HMXBs, and supergiant fast X-ray transients (SFXTs), characterized by short and intense X-ray flares. I discuss the characteristics of these types of supergiant HMXBs, propose a scenario describing the properties of these high-energy sources, and finally show how the observations can constrain the accretion models (e.g. clumpy winds, magneto-centrifugal barrier, transitory accretion disc, etc). Because they are the likely progenitors of Luminous Blue Variables (LBVs), and also of double neutron star systems, related to short/hard gamma-ray bursts, the knowledge of the formation and evolution of this HMXB population is of prime importance.
\end{abstract}


\bodymatter


\section{High-Mass X-ray Binaries}

High energy binary systems are composed of a compact object -- neutron star (NS) or black hole (BH) -- orbiting around and accreting matter from a companion star (see review Ref.~\refcite{chaty:2013}). The companion star is either a low-mass star ($\sim 1 \Msol$ or less, with a spectral type later than B, called LMXB for ``Low-Mass X-ray Binary''), or a luminous early spectral type OB high-mass star ($\sim 10 \Msol$ or more, called HMXB for ``High-Mass X-ray Binary''). $\sim 300$ high energy binary systems are now known in our Galaxy: 187 LMXBs and 114 HMXBs (respectively 62\% and 38\% of the total number, Ref.~\refcite{liu:2007} \& \refcite{liu:2006}). The number of HMXBs is an indicator of star formation rate and starburst activity (Ref.~\refcite{coleiro:2013a}). 
Accretion of matter is different for both types of sources. For LMXBs, the small and low-mass companion star fills and overflows its Roche lobe, and accretion of matter always occurs through the formation of an accretion disc. For HMXBs, while accretion can also occur through an accretion disc for Roche lobe filling systems, this is generally not the case, and there are two alternatives, that we now further describe.

\subsection{Be X-ray binaries}

The first one concerns the case of HMXBs containing a main sequence early spectral type B0-B2e III/IV/V donor star, called in the following BeHMXBs. These rapidly rotating stars possess a circumstellar disc of gas created by a low velocity and high density stellar wind of $\sim 10^{-7} \Msol / yr$. This ``decretion'' disc is characterized by an H$\alpha$ emission line (whose width is correlated with the disc size) and a continuum free-free/free-bound emission, causing an infrared excess (Ref.~\refcite{negueruela:2004}). 
In these systems, accretion periodically occurs, with transient and bright X-ray outbursts: {\it i.} ``Type I'' are regular and periodic outbursts each time the compact object --usually a NS on a wide and eccentric orbit-- crosses the disc at periastron; {\it ii.} ``Type II'' are giant outbursts at any phase, with a dramatic expansion of the disc, enshrouding the NS; {\it iii.} ``Missed'' outbursts exhibit low H$\alpha$ emission (due to a small disc or a centrifugal inhibition of accretion), iv. ``Shifting phase outbursts'' are likely due to the rotation of density structures in the circumstellar disc\footnote{For more details about BeHMXBs, we warmly recommend the excellent review on optical/infrared emission of HMXBs: Ref.~\refcite{charles:2006}.}.
%

\subsection{Supergiant X-ray binaries}

The second one concerns HMXBs with an early spectral type supergiant OB I/II donor star, later called sgHMXBs. These massive stars eject a steady, slow and dense wind, radially outflowing from the equator, and the compact object --usually a NS on a circular orbit-- directly accretes the stellar wind through e.g. Bondy-Hoyle-Littleton process.
We distinguish two groups: Roche lobe overflow and wind-fed systems\footnote{Cyg\,X-1 is the only sgHMXB with both Roche lobe overflow and stellar wind accretion, hosting a confirmed BH, probably a rare product of stellar evolution in X-ray binary systems (Ref.~\refcite{podsiadlowski:2003}).}.
The former group constitutes the classical «bright» sgHMXBs with accreted matter flowing via inner Lagrangian point to the accretion disc, causing a high X-ray luminosity ($L_X \sim 10^{38} \ergs$) during outbursts.
The later group concerns close systems ($P_\mathrm{orb} < 15$\,days) with a low eccentricity, the NS accreting from deep inside the strong steady radiative and highly supersonic stellar wind. These systems exhibit a persistent X-ray emission at regular low-level effect ($L_X \sim 10^{35-36} \ergs$), on which are superimposed large variations on short timescales, due to wind inhomogeneities.
%
During their long term evolution, the orbits of sgHMXBs tend to circularize more rapidly with time, while the rate of mass transfer steadily increases (Ref.~\refcite{kaper:2004}). A milestone in the evolution of these binary systems takes place during the so-called ``common envelope phase''. This phase initiates when the compact object penetrates inside the envelope of the companion star, in a rapidly inward spiralling orbit due to a large loss of orbital angular momentum. This phase has been invoked in Ref.~\refcite{paczynski:1976} to explain how high energy binary systems with very short $P_\mathrm{orb}$ can be formed, while both components of these systems -- large stars at their formation -- would not have been able to fit inside a binary system with such a small orbital separation. This phase, while taken into account in population synthesis models, has never been observed yet, probably because it is short (models predict a maximum duration of common envelope phase of only $\sim 1000$\,years \cite{meurs:1989}) compared to the lifetime of a massive star ($\sim 10^{6-7}$\,years). It is a fundamental ingredient to understand the evolution of high energy binary systems \citep{tauris:2006}.

\section{The {\it INTEGRAL} supergiant revolution} \label{section:INTEGRAL} 
The {\it INTEGRAL} observatory is an ESA satellite launched on 17 October
2002 by a PROTON rocket on an eccentric orbit. It hosts 4
instruments: 2 $\gamma$-ray coded-mask telescopes --imager IBIS
and spectro-imager SPI, observing in the range 10 keV-10 MeV, with
a resolution of $12\amin$ and a field-of-view of $19\adeg$--, a
coded-mask telescope JEM-X (3-100 keV), and an optical telescope
(OMC).
The $\gamma$-ray sky seen by {\it INTEGRAL} is very rich, with 723 sources detected, reported in the $4^{th}$ IBIS/ISGRI soft $\gamma$-ray catalogue \citep{bird:2010}, spanning nearly 7 years of observations in the 17-100 keV domain\footnote{Up-to-date list maintained by J.~Rodriguez \& A.~Bodaghee: {\em http://irfu.cea.fr/Sap/IGR-Sources}}.
Among these sources, there are 185 X-ray binaries (representing 26\% of the whole sample, called ``IGRs'' in the following), 255 Active Galactic Nuclei (35\%), 35 Cataclysmic Variables (5\%), and $\sim 30$ sources of other type (4\%): 15 SNRs, 4 Globular Clusters, 3 Soft $\gamma$-ray Repeaters, 2 $\gamma$-ray bursts, etc. 
215 objects still remain unidentified (30\%).
X-ray binaries are separated into 95 LMXBs and 90 HMXBs, each category representing $\sim 13$\% of IGRs. Among identified HMXBs, there are 24 BeHMXBs and 19 sgHMXBs (resp. 31\% and 24\% of HMXBs).

It is interesting to follow the evolution of the ratio between BeHMXBs and sgHMXBs \cite{chaty:2013}. During the pre-{\it INTEGRAL} era, HMXBs were mostly BeHMXBs. In the catalogue of 110 HMXBs\cite{liu:2000}, there were 52 BeHMXBs and 13 sgHMXBs (respectively 47\% and 12\% of HMXBs). Then, the situation drastically changed with the discovery by {\it INTEGRAL} of 24 sgHMXBs: in the catalogue of 114 HMXBs \cite{liu:2006}, there were 60 BeHMXBs and 37 sgHMXBs (respectively 50\% and 32\% of HMXBs). Therefore, while the BeHMXB/HMXB ratio remained stable, the sgHMXB/HMXB ratio nearly tripled.
The ISGRI energy range ($> 20$\,keV), immune to the absorption that prevented the discovery of intrinsically absorbed sources by earlier soft X-ray telescopes, allowed us to go from a study of individual sgHMXBs (such as GX\,301-2, 4U\,1700-377, Vela\,X-1, etc) to a comprehensive study of the characteristics of a whole population of HMXBs \cite{coleiro:2013a}.

The most important result of {\it INTEGRAL} is the discovery of new X-ray wind-accreting pulsars, mainly sgHMXBs -- concentrated towards tangential directions of Galactic arms, rich in star forming regions --, exhibiting common characteristics which previously had rarely been seen, with longer spin periods and higher absorption, compared to previously known sgHMXBs \cite{bodaghee:2007, chaty:2008a, rahoui:2008a, coleiro:2013b}.

\section{The classical and obscured, the fast and eccentric}

After an extensive multi-wavelength study of IGR sources by various groups, we have reached the consensus that sgHMXBs can be sub-divided into two main classes, which are likely connected, representing two distinct positions within the continuum of their general characteristics:

\subsection{Classical and obscured sgHMXBs}

There are $\sim 16$ classical and persistent sgHMXBs, nearly half of them exhibit a substantial intrinsic and local extinction: there are $\sim 8$ such obscured sgHMXBs, the most extreme example being the highly absorbed source IGR~J16318-4848\citep{chaty:2012a}.
These systems share the following common properties: O8-B1 spectral type companion stars, $\nh \geq 10^{23} \cmmoinsdeux$, compact object on a short/circular orbit ($P_{orb} \sim 3.7-9.7$\,days), and luminous X-ray emission ($L_X = 10^{36-38} \ergs$). In transition to Roche Lobe Overflow, these systems are characterized by slow winds, causing a deep spiral-in, leading to Common Envelope Phase. A dichotomy exists in the orbital period: systems with P$_{orb} \leq 6$\,days\footnote{IGR\,J16393-4611, IGR\,J16418-4532, IGR\,J18027-2016, 4U\,1538-522, 4U\,1700-37, 4U\,1909+07 and XTE\,J1855-026}, and systems with P$_{orb} \geq 9$\,days\footnote{IGR\,J16320-4751, IGR\,J19140+0951, EXO\,1722-363, GX\,301-2, OAO\,1657-415, 1A\,0114+650, 1E\,1145.1-6141 and Vela\,X-1}. 

\subsection{Fast and eccentric sgHMXBs}

Fast and transient X-ray outbursts --an unusual characteristic among HMXBs--, are the signature of the so-called Supergiant Fast X-ray Transients (SFXTs \cite{negueruela:2006a}). There are 17 (+5 candidate) SFXTs (thus representing a significant subclass of sgHMXBs)\footnote{IGR\,J08408-4503, IGR\,J11215-5952, IGR\,J16195-4945, IGR\,J16207-512, IGR\,J16328-4726, IGR\,J16418-4532, IGR\,J16465-4507, IGR\,J16479-4514, IGR\,J17354-3255, IGR\,J17544-2619 (their archetype \citep{pellizza:2006}), IGR\,J18462-0223, IGR\,J18483-0311, SAX\,J1818.6-1703, XTE\,J1739-302, AX\,J1820.5-1434, AX\,J1841.0-0536 and AX\,J1845.0-0433}. These SFXTs are divided in three sub-classes: 7 classic-like systems, 4 fast transients reaching anomalously low luminosities, 3 eccentric systems (and 3 unclear sources).
Their common characteristics are: a compact object on a short/circular orbit, transient and intense X-ray flares detected any time on the orbit, rising in tens of minutes ($L_X \sim 10^{35-37} \ergs$), lasting a few hours, and alternating with long ($\sim 70$\,days) quiescence ($L_X \sim 10^{32-34} \ergs$), with an impressive variability factor $\frac{L_{max}}{L_{min}}$ going from $15-50$ to $10^{2-5}$. They host OB supergiant companions, with P$_{orb} \sim 3.3 - 54$\,days; absorbed ($\nh \sim 10^{22} \cmmoinsdeux$) cut-off (10-30 keV) power-law X-ray spectrum; with X-ray pulsations (from a few to 1000s seconds) implying that they are young neutron star ($B \sim 10^{11-12}$\,G).

\subsection{Accretion processes}

Various accretion processes have been proposed to account for X-ray properties of different kinds of sgHMXBs, we enumerate here the main mechanisms\footnote{All references are in Ref.~\refcite{chaty:2013}.}: 
i.)~Clumpy stellar wind accretion, 
ii.)~Magnetic/centrifugal gating mechanism, 
iii.)~Hydrodynamic properties of accretion stream («breathing» of shock front), 
iv.)~Formation/dissipation of temporary accretion disc, 
and v.)~Cooling switch. 

\section{Conclusions} \label{section:conclusion}

While the {\it INTEGRAL} satellite was not primarily designed for this, it allowed a great progress in the study of HMXBs in general, and of sgHMXBs in particular. Let us recall the {\it INTEGRAL} legacy on sgHMXBs.
First, the {\it INTEGRAL} satellite tripled the total number of known sgHMXBs in our Galaxy, most of them being slow and absorbed wind-fed accreting X-ray pulsars.
Second, the {\it INTEGRAL} satellite revealed the existence in our Galaxy of previously hidden populations of high energy binary systems: i.) a population of persistent and obscured sgHMXBs, exhibiting long $P_\mathrm{spin}$ ($\sim1$\,ks) and strong intrinsic absorption (large $\nh$, with the NS deeply embedded in the dense stellar wind).
ii.) the SFXTs, hosting supergiant companion stars and exhibiting brief and intense X-ray flares (luminosity $L_X \sim 10^{36} \ergs$ at the peak, during a few ks every $\sim 7$\,days), which can be explained by various accretion processes.

\section*{Acknowledgments}

I would like to warmly thank Thomas Tauris, as an efficient organizer and chairman of the parallel session BN3 - Double Neutron Stars and Neutron Star-White Dwarf Binaries. I am lifelong indebted to my close collaborators on the study of {\it INTEGRAL} sources: A. Coleiro, P.A. Curran, Q.Z. Liu, I. Negueruela, L. Pellizza, F. Rahoui, J. Rodriguez, M. Servillat, J.A. Tomsick, J.Z. Yan, and J.A. Zurita Heras. This work, supported by the Centre National d'Etudes Spatiales (CNES), was based on observations obtained with MINE --Multi-wavelength {\it INTEGRAL} NEtwork--.


%

\end{document}